\documentclass[floats,aps,showpacs,twocolumn,superscriptaddress]{revtex4}

\usepackage[dvips]{graphicx}
\usepackage{amsfonts}
\usepackage{amsmath,amssymb}

\newcommand{\kF}{k_{\text{F}}}
\newcommand{\pF}{p_{\text{F}}}
\newcommand{\EF}{E_{\text{F}}}
\newcommand{\heff}{h_{\text{eff}}}
\newcommand{\Areg}{A_{\text{reg}}}
\newcommand{\areg}{a_{\text{reg}}}
\newcommand{\APS}{A_{\text{PS}}}
\newcommand{\hmin}{h_\text{min}}
\newcommand{\cA}{{\cal A}}
\newcommand{\Int}{\int\limits}
\newcommand{\ud}{\text{d}}
\newcommand{\aiz}{{z_0}}
\DeclareMathOperator{\Ai}{Ai}

\begin{document}

\title{Nano--wires with surface disorder:\\ 
Giant localization lengths and quantum--to--classical crossover}

\author{J.~Feist}
\email{johannes.feist@tuwien.ac.at}
\affiliation{Institute for Theoretical Physics, 
             Vienna University of Technology, 1040 Vienna, Austria}

\author{A.~B\"acker}
\affiliation{Institut f\"ur Theoretische Physik, Technische Universit\"at
             Dresden, 01062 Dresden, Germany}

\author{R.~Ketzmerick}
\affiliation{Institut f\"ur Theoretische Physik, Technische Universit\"at
             Dresden, 01062 Dresden, Germany}

\author{S.~Rotter}
\affiliation{Institute for Theoretical Physics, 
             Vienna University of Technology, 1040 Vienna, Austria}
\affiliation{Department of Applied Physics, Yale University, New Haven, Connecticut, 06520, USA}

\author{B.~Huckestein}
\affiliation{Institut f\"ur Theoretische Physik III, Ruhr--Universit\"at Bochum,
             44780 Bochum, Germany}

\author{J.~Burgd\"orfer}
\affiliation{Institute for Theoretical Physics, 
             Vienna University of Technology, 1040 Vienna, Austria}

\date{June 20, 2006}

\begin{abstract}
  We investigate electronic quantum transport through nano--wires with
  one--sided surface roughness.  A magnetic field perpendicular to the
  scattering region is shown to lead to exponentially diverging localization
  lengths in the quantum--to--classical crossover regime.  This effect can be
  quantitatively accounted for by tunneling between the regular and the
  chaotic components of the underlying mixed classical phase space.
\end{abstract}
\pacs{05.45.Mt, 72.20.Dp, 73.23.Ad, 73.63.Nm}

\maketitle

\noindent

Transport through a disordered medium is a key issue in solid state physics
which comprises countless applications in (micro--) electronics and
optics~\cite{sheng95LeeRam1985altshuler91}.  
The ubiquitous presence of
disorder plays a prominent role for the behavior of transport coefficients
governing, e.g., the metal--insulator transition~\cite{anderson58}.  The
interest in disordered media has recently witnessed a revival due to new
experimental possibilities to study the 'mesoscopic' regime of transport where
a quantum--to--classical crossover gives rise to a host of interesting
phenomena~\cite{datt95ferr97}.

In most investigations a static disorder is assumed to be present in the
\emph{bulk} of a material.  The strength and distribution of the disorder
potential determine whether transport will be ballistic, diffusive, or
suppressed in the localization
regime~\cite{sheng95LeeRam1985altshuler91,datt95ferr97}.
In nano--devices the reduction of system sizes leads, however, to an
increased surface--to--volume ratio, for which \emph{surface roughness} can
represent the dominant source of disorder scattering. While random matrix
theory (RMT) is successful in describing bulk disordered
systems~\cite{Bee1997}, its application to wires with surface disorder is not
straightforward~\cite{SurfDis}.

\begin{figure}[!b]
\begin{center}
\includegraphics[angle = 0, width = 85mm]{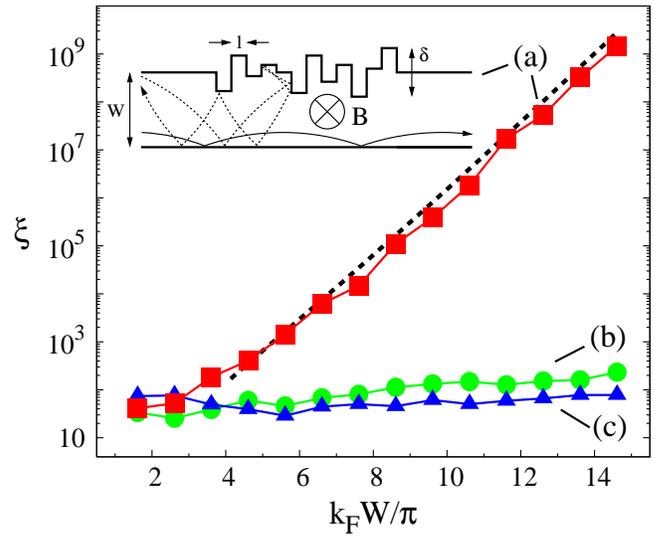}
\caption{(Color online) 
  Localization length $\xi$ for a wire
  with surface roughness vs 
  $\kF W/\pi=1/\heff$. 
  Results are compared for wires with 
  (a) one--sided disorder (OSD) with $B \neq 0$
  (red ${\scriptstyle\blacksquare}$), 
  (b) OSD with $B=0$ (green {\large $\bullet$}), and 
  (c) two--sided disorder
  with $B \neq 0$ (blue $\blacktriangle$). 
  In (a) an exponential increase
  of $\xi$ is observed in excellent agreement with Eq.~(\ref{eq:final})
  which has no adjustable parameters  (dashed line).}
\label{fig:1}
\end{center}
\end{figure}

In the present paper we study electronic quantum transport through a
nano--wire in the presence of one--sided surface disorder and a magnetic
field. We show both numerically and analytically that by increasing
the number of open channels $N$ in the wire, or equivalently, by increasing
the wavenumber $\kF$, the localization length $\xi$ increases exponentially.  
Using a numerical approach that allows to study extremely long wires 
we show an increase by a factor $10^7$ (Fig.~\ref{fig:1}).  
Such a giant localization length falls outside the scope
of RMT predictions, $\xi\propto N$, previously
studied for this system~\cite{GarGovWoe2002}. Instead it 
can be understood in terms of the underlying mixed
regular--chaotic classical motion in the wire.  We find that the
conductance through the wire is controlled by tunneling from the regular
to the chaotic part of phase space. This process, often referred to as
``dynamical tunneling''~\cite{DavHel81}, has been actively studied in quantum
chaos and plays an important role in the context of classically
transporting phase space
structures~\cite{HanOttAnt1984,FisGuaReb2002SchDitKet2005,HufKetOttSch2002IomFisZas2002,BaeKetMon2005,PruSch2006}.
Here we establish a direct quantitative link between the exponential increase
of the localization length in mesoscopic systems and the suppression of
tunneling from the regular to the chaotic part of phase space in the
semiclassical limit.

We consider a 2D wire with surface disorder to which two leads
of width $W$ are attached (Fig.~\ref{fig:1}, inset), with a homogeneous
magnetic field $B$ perpendicular to the wire present throughout the system.
We simulate the disorder by a random sequence of vertical steps.
The wire can thus be assembled from rectangular elements, referred to
in the following as modules, with equal width $l$, but random heights $h$,
uniformly distributed in the interval $[W-\delta/2,W+\delta/2]$.  This
particular representation of disorder allows for an efficient numerical 
computation of quantum transport 
for remarkably long wires $L\to\infty$ by employing the modular recursive
Green's function method~\cite{RotWeiRohBur2003}.  
We first
calculate the Green's functions for $M\!=\!20$ rectangular modules with 
different heights.  
A random sequence of these modules is connected by means of a matrix
Dyson equation.  Extremely long wires can be reached by implementing an
``exponentiation'' algorithm~\cite{SkjHauSch1994}: Instead of connecting the
modules individually, we iteratively construct different generations of
``supermodules'', each consisting of a randomly permuted sequence of $M$
modules of the previous generation. Repeating this process leads to the
construction of wires whose length increases exponentially with the number of
generations 
\footnote{With this approach we can study wires with up to $\sim
10^{12}$ modules, beyond which numerical unitarity deficiencies set
in.  For wires with up to $10^5$ modules we can compare this supermodule
technique containing pseudo--random sequences with truly random sequences of
modules.  For configuration--averaged transport quantities the results are
indistinguishable from each other.}.
 
The transmission $(t_{mn})$ and reflection amplitudes $(r_{mn})$ for an electron
injected from the left are evaluated by projecting the Green's function at the
Fermi energy $\EF$ onto all lead modes $m,n\in \{1,\ldots,N \}$ in the
entrance and exit lead, respectively.  Here $N=\lfloor \kF W/\pi \rfloor$ is
the number of open lead modes and $\kF$ the Fermi wave number.  We obtain the
localization length $\xi$ in a wire composed of $L$ modules (i.e.\ length
$Ll$) by analyzing the dimensionless conductance $g={\rm Tr} (t^{\dagger} t)$
in the regime $g \ll 1$, extracting $\xi$ from $\langle {\rm
  ln}\,g\rangle\sim -L/\xi$.  The brackets $\langle\ldots\rangle$ indicate
the ensemble average over 20 different realizations of disorder and 3
neighboring values of wave numbers $\kF$.

For increasing $\kF$ we adjust 
the magnetic field $B$ such that the cyclotron radius $r_c= \hbar
\kF/ (eB)$ remains constant.
This leaves the classical dynamics invariant and allows for
probing the quantum--to--classical crossover as 
$\kF \to\infty$. We choose  $r_c=3W$ and a disorder amplitude $\delta=(2/3)W$
such that we obtain a large regular region in phase space (see below)
and use a module width
$l=W/5$.
We find for one--sided disorder  an 
exponential increase of the localization length $\xi$ (Fig.~\ref{fig:1}),
while $\xi$ remains almost constant when either (i)
the magnetic field is switched off or (ii) a two--sided disorder is
considered. The latter clearly rules out that the observed giant localization
length is due to edge states of the quantum Hall effect~\cite{datt95ferr97}.

Before giving an analytic determination of the exponentially increasing
localization length, we provide an explanation invoking the
mixed classical phase--space structure which captures the essential features of
this increase.

\begin{figure}[tb]
  \begin{center}
  \includegraphics[angle = 0,width = 86mm]{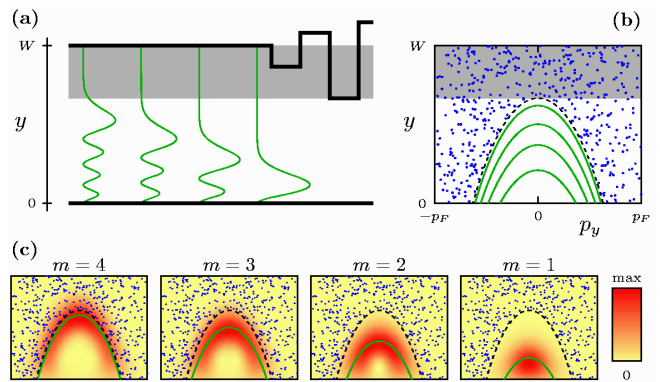}
    \caption{(Color online) 
      (a) Nano--wire with the regular transverse modes (green) 
      $m=4, 3, 2, 1$ for $\kF W/\pi=14.6$.  The gray shaded part
      indicates the $y$-range affected by disorder.  (b) Poincar\'e section
      showing a large regular island with outermost torus (dashed), a
      chaotic sea (blue dots) and quantized tori corresponding to the regular
      modes (green).  (c) Poincar\'e--Husimi functions of these modes
      and their quantizing tori. 
      \label{fig:section}}
  \end{center}
\end{figure}

The classical dynamics inside the disordered wire is displayed by a Poincar\'e
section in Fig.~\ref{fig:section}b, for a vertical cut at the wire
entrance ($x=0$) with periodic boundary conditions in the $x$-direction.
The resulting section ($y$, $p_y$) for $p_x> 0$ 
shows a large regular region with invariant
curves corresponding to skipping motion along the lower straight boundary of
the wire. Close to the upper disordered boundary ($y>W-\delta/2$) the motion 
appears to be chaotic for all $p_y$. 
A corresponding Poincar\'e section for
$p_x<0$ (not shown) is globally chaotic. 
The lowest transverse modes (Fig.~\ref{fig:section}a)
of the incoming scattering wave functions
overlap primarily with the regular island
(Fig.~\ref{fig:section}b). Only their exponential tunneling tail through the 
diamagnetic potential barrier (in Landau gauge)
\begin{equation}\label{eq:diamagpot}
  V(y)=\frac{1}{2} m_e \omega_c^2 (y-y^0)^2 - \EF
\end{equation}
touches the upper disordered surface at $y>W-\delta/2$. In
Eq.~(\ref{eq:diamagpot}), $m_e$
is the electron mass, $\omega_c$ the cyclotron frequency, and $y^0$ the
guiding center coordinate. These regular modes can be semiclassically
quantized as~\cite{KosLif1955,BeeHou1988}
\begin{equation} \label{eq:WKB}
  \frac{A}{h} = \frac{B \cA}{h/e} = (m-1/4) \quad \text{with $m=1,2,...$} \;\;,
\end{equation}
where $A$ is the area in the Poincar\'e section enclosed by a quantized torus
and $\cA=r_c A /\pF$ is the area in position space enclosed by a segment of a
skipping orbit.  One finds $A=\pF r_c \left[\arccos (1-\nu) - (1-\nu)
  \sqrt{1-(1-\nu)^2}\right]$ for $0 \leq \nu \leq \nu_{\text{max}} \leq 1$,
where $\nu r_c$ is the $y$-position at the top of the cyclotron orbit.  The
size $\Areg$ of the regular island is found for
$\nu=\nu_{\text{max}}=(W-\delta/2)/r_c$. The Poincar\'e--Husimi projections
(i.e.~projections onto coherent states of the transverse eigenfunctions show,
indeed, a density concentration near the quantized tori residing in the
regular region of phase space (Fig.~\ref{fig:section}c). 

The lowest mode $m=1$ in the center of the island has the 
smallest tunneling rate 
\cite{HanOttAnt1984, PodNar2003SchEltUll2005:pShe2005, SheFisGuaReb2006}
\begin{equation} \label{eq:gammaestimate}
  \gamma_1 \sim \exp\left(- C \, \frac{\Areg}{h} \right)
\end{equation}
to the chaotic sea with some constant $C$ (see below).
Its temporal decay $\exp(-\gamma_1 t)$
together with its velocity $v_1=\hbar k_x/m_e$
lead to an exponential decay as a
function of propagation length $x$, $\exp(-\gamma_1 x/v_1)$.
This gives a localization length 
$\xi \sim \gamma_1^{-1}$~\cite{HufKetOttSch2002IomFisZas2002}.
When increasing $\kF$, while keeping the
cyclotron radius $r_c$ fixed, the classical dynamics remains invariant while
the island area scales as $\Areg=\areg \APS$.
Here $\APS = 2 \pF W$
is the area of the Poincar\'e section and $\areg$ is the relative
size of the island.
This semiclassical limit is thus equivalent to
decreasing the effective Planck's constant $\heff:=h/\APS = (\kF W/\pi)^{-1}$ 
and results in an
exponential increase of the localization length
\begin{equation}\label{eq:xiestimate}
  \xi \sim  \exp\left(C \, \frac{\areg}{\heff}\right),
\end{equation}
for $\heff\to0$,
qualitatively explaining Fig.~\ref{fig:1}a. 
Moreover,
this exponential increase should set in when the first mode fits into
the island, i.e.\ for $\Areg/h \approx 1$. For the parameters of
Fig.~\ref{fig:1}a we have $\nu_{\text{max}}=2/9$, resulting in the critical
value $\kF W/\pi \approx 3.5$ which is in very good agreement with the
numerical result.  By contrast, for two-sided disorder or for $B=0$ no regular
island with skipping orbits exists and $\xi$ shows no
exponential increase, see Fig.~\ref{fig:1}.

We now turn to an analytical derivation of the localization length 
using the specifics of the scattering geometry (Fig.~\ref{fig:1} inset). 
To this end we first
calculate the transmission amplitude $t_{11}$ of the 
transverse regular mode $m=1$ by considering
its consecutive projections from one module to the next
\begin{equation}\label{eq:transamplitude}
   t_{11} \approx \prod_{j=1}^{L-1}
            \Int_{0}^{W+\delta/2} 
            \chi_{h(j)}(y) \;
            \chi_{h(j+1)}(y) \; \ud y \,,
\end{equation}
where $\chi_{h(j)}(y)$ is the mode wave function  in module $j$ with
height $h(j)$.  Eq.~(\ref{eq:transamplitude}) amounts to a sequence of \emph{sudden
approximations} for the transition amplitude between adjacent surface steps.
As the wave function is exponentially suppressed at the upper boundary, the
scale $l\kF$ introduced by the corners drops out of the calculation.
For simplicity, a few technical approximations have been invoked, whose
accuracy can be checked numerically: i) terms in the transmission from one
module to the next that involve reflection coefficients and are typically
smaller by a factor of 5 are neglected, ii) contributions from
direct coupling between different island modes are neglected, and iii) the factor
$(2y\!-\!y_{h(j)}^0\!-\!y_{h(j+1)}^0)$ from the orthonormality relation for
the $\chi$ functions~\cite{RotWeiRohBur2003} is omitted in the above integral as its
contribution is negligible.

The modes pertaining to different heights $h$ can be written as 
$\chi_h(y)=[\chi_\infty(y) - \varepsilon_h(y)]/N_h$, 
where $\chi_\infty(y)$ is the mode wave function if there
was no upper boundary, $\varepsilon_h(y)$ is the correction that is largest at
the upper boundary (where $\chi_h(h)$=0), and $N_h$ is a normalization
factor. Keeping only terms of order $O(\varepsilon_h)$
and using a WKB approximation for $\varepsilon_h(y)$ around $y=\hmin=W-\delta/2$ leads to
\begin{equation} \label{eq:t11}
  t_{11} \approx (1-\sigma)^{2L/M}
  \text{ with }
  \sigma = \frac{\chi^{2}_{\infty}(\hmin)}{\kF \sqrt{V(\hmin)/\EF}}\,.
\end{equation}
According to Eq.~(\ref{eq:t11})
the coupling strength is quantitatively determined by the tunneling electron
density at $y=\hmin$ in the classically forbidden region
of the 1D diamagnetic potential, Eq.~(\ref{eq:diamagpot}).
The conductance in the regime $g \ll 1$ is now given by
\begin{equation}
    g=|t_{11}|^2\approx\exp(-4\sigma L/M)\,,
\end{equation}
resulting in a localization length $\xi=M/(4\sigma)$.
Using a WKB approximation for $\chi_\infty(y)$ we find
\begin{equation}\label{eq:final}
  \xi(\heff)  \approx \left(a \, \heff^{-2/3} - b \right) 
        \exp \left(c \, \heff^{-1} (1-d\, \heff^{2/3})^{3/2} \right)\,,
\end{equation}
with coefficients $a=(16\pi^{5})^{1/3}\zeta M \eta \rho^{-1/3}$,
$b = -2 \pi \aiz \zeta M$,
$c = \pi (32/9)^{1/2} \eta^{3/2} \rho^{-1/2} [1 \!+\! (3/20)\eta \rho^{-1}]$,
$d = - \aiz \rho^{1/3}/(2^{1/3}\pi^{2/3} \eta)$. Here $\aiz\approx -2.338$ is the 
first zero of the Airy function $\Ai(z)$, $\zeta = \int_{\aiz}^{\infty} \Ai(z)^2 \ud z$, 
$\eta=\hmin/W$, and $\rho=r_c/W$ are dimensionless parameters 
\footnote{Setting $d=0$ corresponds to a quantization
at the minimum of the diamagnetic potential $V(y)$ in Eq.~(\ref{eq:diamagpot}).
This produces the correct leading order, but for our
largest $1/\heff=14.6$ the result would be wrong by a factor $10^5$.}.
Eq.~(\ref{eq:final})
is in very good quantitative agreement with the numerically determined
localization length (Fig.~\ref{fig:1}a). We conclude that tunneling 
from the regular phase space island is primarily due 
to interaction of each regular mode with the rough surface rather than 
via successive transitions from inner to outer island modes.

We note that the constant $C$ 
in Eqs.~(\ref{eq:gammaestimate}) and (\ref{eq:xiestimate})
is found to be $C=2\pi
[1+(289/960)\eta\rho^{-1}]$, which differs from
$C=2\pi$~\cite{SheFisGuaReb2006} and $C=3-\ln
4$~\cite{PodNar2003SchEltUll2005:pShe2005} derived for other examples
of dynamical tunneling from a resonance-free regular island to a chaotic sea.
We also note that the scaling behavior of $\xi$ in Eq.~(\ref{eq:final}) is reminiscent 
of previously obtained results for diffusive 2D systems (see \cite{sheng95LeeRam1985altshuler91}).

\begin{figure}[bt]
  \begin{center}
  \includegraphics[angle = 0,width = 86mm]{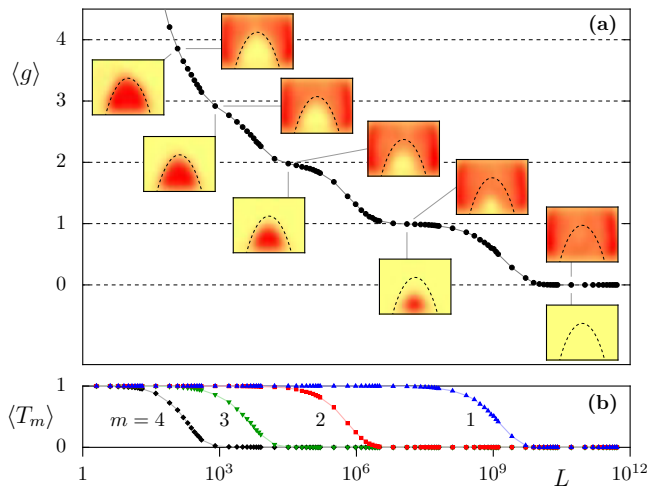}
    \caption{(Color online) (a) Averaged conductance $\langle g \rangle$
      vs~length $L$ of the wire.  
The step-wise decrease is accompanied by the disappearance
             of the regular modes and the flooding of the island 
             region by chaotic states. The
             Poincar\'e--Husimi distributions
             to the left (right) of the curve correspond to scattering
from left to right (backscattering from right to right).
(b) Transmission $\langle T_m
      \rangle$ of the incoming mode $m$ vs~$L$.
      \label{fig:3}}
  \end{center}
\end{figure}

For the case of a constant magnetic field $B$,
increasing $\kF$ increases
the cyclotron radius, $r_c \propto \kF$, and
the classical dynamics is no longer invariant. In particular the area of the
regular island $\Areg \sim \sqrt{\kF}$ shrinks compared to 
$\APS\sim\kF$ as
skipping motion is increasingly suppressed. Nevertheless, the arguments
leading to Eqs.~(\ref{eq:xiestimate}) and~(\ref{eq:final}) remain
applicable and yield a localization length that increases dramatically as $\xi
\sim \exp(\text{const} \, \sqrt{\kF})$ in agreement with numerical
observations (not shown).

Now we turn to the behavior of the conductance for wires of lengths smaller
than the localization length. Modes with larger $m$ have larger amplitudes
near the rough surface and thus couple more strongly to the chaotic part of
phase space.  They have, consequently, larger tunneling rates $\gamma_m$ and
smaller localization lengths $\xi_m \sim \gamma_m^{-1}$. The successive
elimination of modes as a function of the length $L$ of the wire
results in a sequence of plateaus (Fig.~\ref{fig:3}a).
For $L>\xi_m$ the mode
$m$ no longer contributes to transport, as can be seen by its individual
contribution to the transmission in Fig.~\ref{fig:3}b.
This disappearance of regular modes 
is reflected in the
averaged Poincar\'e--Husimi distributions calculated
from incoherent superpositions 
of all modes entering from the left and scattering to the right.
Also shown are the complementary distributions obtained for 
backscattering from right to right. For small $L$
these Poincar\'e--Husimi functions are outside
the regular island, while with increasing $L$ they begin to ``flood'' 
it~\cite{BaeKetMon2005}. 
This process is complete for lengths $L\gg\xi$.
The complementarity of the Husimi distributions
illustrates that tunneling between the regular
island and the chaotic sea proceeds symmetrically in both directions, as required by the
unitarity of the scattering matrix.

Summarizing, we have presented a numerical computation and an analytical
derivation for the exponential increase of the localization length in a
two-dimensional system of a quantum wire with one--sided surface disorder.
Our approach, based on a mixed phase-space analysis,
also explains the increase of $\xi$ over one order of magnitude
under increase of the magnetic field observed in Ref.~\cite{GarGovWoe2002}.
It sets in for a magnetic field for which the regular island is large enough to 
accomodate at least one quantum mechanical mode.
Clearly, the RMT result, $\xi\propto N$,
which ignores the mixed phase-space structure, no longer applies.
Instead, we find that the giant localization length (Fig.~\ref{fig:1})
in this disordered mesoscopic device is
determined by the tunneling from the regular to the chaotic region, the rate of
which is exponentially suppressed in the semiclassical regime.

AB and RK acknowledge support by the DFG under
Contract No.~KE 537/3-2, JF, SR and JB acknowledge support by the FWF--Austria
(Grant P17354) and the Max-Kade foundation, New York.


\begin{thebibliography}{10}

\bibitem{sheng95LeeRam1985altshuler91}
P. Sheng, {\em Introduction to Wave Scattering, Localization and Mesoscopic
  Phenomena} (Academic, New York, 1995); P.~A. Lee and T.~V. Ramakrishnan, Rev.
  Mod. Phys. {\bf 57}, 287 (1985); {\em Mesoscopic Phenomena in Solids}, edited
  by B.~L. Altshuler, P.~A. Lee, and R.~A. Webb (North-Holland, Amsterdam,
  1991).

\bibitem{anderson58}
P.~W. Anderson, Phys. Rev. {\bf 109},  1492  (1958).

\bibitem{datt95ferr97}
S. Datta, {\em Electronic Transport in Mesoscopic Systems} (Cambridge
  University Press, Cambridge, 1995); D.~K. Ferry and S.~M. Goodnick, {\em
  Transport in Nanostructures} (Cambridge University Press, Cambridge, 1997).

\bibitem{Bee1997}
C.~W.~J. Beenakker, Rev. Mod. Phys. {\bf 69},  731  (1997).

\bibitem{SurfDis}
M. Leadbeater, V.~I. Falko, and C.~J. Lambert, Phys. Rev. Lett. {\bf 81}, 1274
  (1998); J.~A. S\'anchez-Gil, V. Freilikher, I. Yurkevich, and A.~A.
  Maradudin, Phys. Rev. Lett. {\bf 80}, 948 (1998); A. Garc\'ia-Mart\'in and
  J.~J. Saenz, Phys. Rev. Lett. {\bf 87}, 116603 (2001); F.~M. Izrailev, J.~A.
  Mendez-Bermudez, and G.~A. Luna-Acosta, Phys. Rev. E {\bf 68}, 066201 (2003);
  E.~I. Chaikina, S. Stepanov, A.~G. Navarrete, E.~R. Mendez, and T.~A.
  Leskova, Phys. Rev. B {\bf 71}, 085419 (2005).

\bibitem{GarGovWoe2002}
A. Garc\'ia-Mart\'in, M. Governale, and P. W\"olfle, Phys. Rev. B {\bf 66},
  233307  (2002).

\bibitem{DavHel81}
M.~J. Davis and E.~J. Heller, J. Chem. Phys. {\bf 75},  246  (1981).

\bibitem{HanOttAnt1984}
J.~D. Hanson, E. Ott, and T.~M. Antonsen, Phys. Rev. A {\bf 29},  819  (1984).

\bibitem{FisGuaReb2002SchDitKet2005}
S. Fishman, I. Guarneri, and L. Rebuzzini, Phys. Rev. Lett. {\bf 89}, 084101
  (2002); H. Schanz, T. Dittrich, and R. Ketzmerick, Phys. Rev. E {\bf 71},
  026228 (2005).

\bibitem{HufKetOttSch2002IomFisZas2002}
L. Hufnagel, R. Ketzmerick, M.-F. Otto, and H. Schanz, Phys. Rev. Lett. {\bf
  89}, 154101 (2002); A. Iomin, S. Fishman, and G.~M. Zaslavsky, Phys. Rev. E
  {\bf 65}, 036215 (2002).

\bibitem{BaeKetMon2005}
A. B\"acker, R. Ketzmerick, and A.~G. Monastra, Phys. Rev. Lett. {\bf 94},
  054102  (2005).

\bibitem{PruSch2006}
M. Prusty and H. Schanz, Phys. Rev. Lett. {\bf 96},  130601  (2006).

\bibitem{RotWeiRohBur2003}
S. Rotter {\it et al.}, Phys. Rev. B {\bf 62}, 1950 (2000);
{\bf 68}, 165302 (2003).

\bibitem{SkjHauSch1994}
J. Skj\r{a}nes, E.~H. Hauge, and G. Sch\"on, Phys. Rev. B {\bf 50},  8636
  (1994).

\bibitem{KosLif1955}
A.~M. Kosevich and I.~M. Lifshitz, Zh. Eksp. Teor. Fiz. {\bf 29},  743  (1955),
  [Sov. Phys. JETP {\bf 2} 646 (1956)].

\bibitem{BeeHou1988}
C.~W.~J. Beenakker and H. van Houten, Phys. Rev. Lett. {\bf 60},  2406  (1988).

\bibitem{PodNar2003SchEltUll2005:pShe2005}
V.~A. Podolskiy and E.~E. Narimanov, Phys. Rev. Lett. {\bf 91}, 263601 (2003);
  P. Schlagheck, C. Eltschka, and D. Ullmo, arXiv.org:nlin.CD/0508024 (2005); M.
  Sheinman, Master thesis, Technion (2005).

\bibitem{SheFisGuaReb2006}
M. Sheinman, S. Fishman, I. Guarneri, and L. Rebuzzini, Phys. Rev. A {\bf 73},
  052110  (2006).

\end{thebibliography}
\end{document}